\documentclass{article}
\usepackage[a4paper,margin=1in]{geometry}
\usepackage[margin=1in]{geometry}
\usepackage{amsmath,amssymb}
\usepackage{graphicx}
\usepackage{bm}
\usepackage{siunitx}
\usepackage{hyperref}
\usepackage{xcolor}
\usepackage{cite}
\usepackage{setspace}
\usepackage{newunicodechar}
\newunicodechar{−}{-}

\title{\bf Enhanced Electron--Phonon Coupling and Superconductivity in Ba-Alloyed A15 LaH$_{5.75}$}

\author{
Jakkapat Seeyangnok$^{1}$,
Udomsilp Pinsook$^{1}$,
Graeme J Ackland$^{2}$\\[1ex]
$^{1}$Department of Physics, Faculty of Science, \\
Chulalongkorn University, Bangkok, Thailand\\
$^{2}$Centre for Science at Extreme Conditions,\\
School of Physics and Astronomy,\\
University of Edinburgh, Edinburgh, United Kingdom\\[1ex]
\texttt{jakkapat.se@chula.ac.th}\\
}

\date{} 

\begin{document}

\maketitle

\begin{abstract}
Recent experiments have established rare-earth A15-type hydrides as a distinct family of high-temperature superconductors that can be stabilized at significantly lower pressures relative to other superconducting hydrides. In particular, 
A15-type LaH$_{5.75}$, was recently shown to be a high T$_c$ superconductor.
We have investigated a range of ternary substitutions and  demonstrate that partial substitution of La by Ba stabilizes the A15 hydride lattice, with La$_{0.75}$Ba$_{0.25}$H$_{5.75}$ as a stable compound. We calculate a superconducting transition temperature of approximately 183~K, almost double the value for LaH$_{5.75}$ at similar pressures. Ba is typically a 2$^+$ ion while La is 3$^+$.  The reduced number of electrons disrupts the formation of H$_2$ units and shifts the Fermi surface, leading to strongly enhance electron-phonon coupling.  By contrast, addition of Hf fails to produce any stable compunds.
This work extends the emerging alloy-design principles  in A15 hydride superconductors, demonstating that doping can shift the Fermi level and tune the strength of the electron-phonon coupling.  This offers concrete guidance for the experimental realization of new high-$T_c$ phases.
\end{abstract}


\section{Introduction}
The pursuit of high-temperature superconductivity in hydrogen-rich materials has been continuously explored since the discovery of the very first superconductivity in mercury~\cite{onnes1911superconductivity}. In most cases, the BCS mechanism involves phonon-mediated electron-electron interactions at the  Fermi level \cite{frohlich1950theory,bardeen1957theory,migdal1958interaction}. This theory, combined with phonon calculations in density functional theory, enable the calculation of the superconducting phase transition critical temperature $T_c$, where the system evolves from a normal metallic state into a superconducting state characterized by the formation of Cooper pairs as coherent quasiparticles~\cite{nambu1960quasi}.

Achieving a high  $T_c$ in conventional, phonon-mediated superconductors requires a favorable combination of a large electronic density of states at the Fermi level, high characteristic phonon frequencies, and strong electron–phonon matrix elements~\cite{allen1983theory}. Owing to its extremely low atomic mass, hydrogen naturally supports very high-frequency lattice vibrations, making it an ideal candidate for high-$T_c$ superconductivity. Indeed, as long ago as 1968, Ashcroft predicted that solid hydrogen would undergo metallization under extreme compression~\cite{ashcroft1968metallic} and subsequently exhibit superconductivity. However, such pressures are required primarily to suppress the formation of molecular H$_2$ units and stabilize an atomic hydrogen lattice capable of supporting Cooper pairing. A practical strategy to lower the required external pressure is to incorporate hydrogen into compounds with heavier elements, whose electrons disrupt covalent H$_2$ bonding and stabilize dense hydrogen networks pressures~\cite{ashcroft2004hydrogen,mcmahon2011high}. This  approach has led to the discovery of a wide range of hydrogen-rich binary and ternary hydrides exhibiting superconductivity at experimentally accessible pressures.

This approach has led to the discovery of record-breaking superconductors, such as H$_3$S with a superconducting transition temperature exceeding 200~K at pressures around 150~GPa~\cite{duan2014pressure,drozdov2015conventional,huang2019high,einaga2016crystal,capitani2017spectroscopic} and LaH$_{10}$ with $T_c$ values approaching 250~K at approximately 170~GPa~\cite{somayazulu2019evidence,drozdov2019superconductivity}. In these systems, dense hydrogen sublattices and strong electron--phonon coupling are the key factors driving the exceptionally high superconducting transition temperatures. Building on these successful discoveries, the search for new binary hydride phonon-mediated high-temperature superconductors has been pursued intensively~\cite{sun2024clathrate,gao2021superconducting}. This effort has led to a number of experimental realizations, including ThH$_{10}$~\cite{semenok2020superconductivity}, YH$_6$~\cite{troyan2021anomalous}, YH$_9$~\cite{kong2021superconductivity}, CeH$_9$ and CeH$_{10}$~\cite{chen2021high}, and CaH$_6$~\cite{ma2022high,li2022superconductivity}. In addition, extending alloying strategies to multicomponent metal sublattices has been shown to play an important role in enhancing superconductivity in ternary hydrides~\cite{zhao2024superconducting,wrona2025high}. This approach has stimulated numerous theoretical predictions of new superconducting systems~\cite{zhu2025predictions,shutov2024ternary,yan2024ternary,jiang2025data,ma2017prediction,wu2024superconducting}, including ternary superconducting hydrides adopting the A15-type crystal structure~\cite{wei2023designing}. 

In this context, A15-type hydrides have recently emerged as a promising class of superconductors bridging this gap, with a variety of rare-earth A15 hydrides now experimentally synthesized. Representative examples include LaH$_{5.75}$~\cite{lah575_prl,cross2024high,guo2024unusual} with $T_c \approx 100$~K near 100~GPa, EuH$_{5.75}$~\cite{semenok2020novel} with $T_c \approx 100$~K at around 100~GPa, LuH$_{5.75}$~\cite{li2023superconductivity} exhibiting $T_c \approx 71$~K at 218~GPa, and YH$_{5.75}$~\cite{siska2024ultrafast,aslandukova2024diverse} with $T_c \approx 100$~K near 100~GPa. Beyond binary hydrides, chemical alloying has recently been established as an effective strategy for enhancing superconductivity in A15 systems. A notable example is the experimentally synthesized ternary A15-(Lu,Y)H$_{5.75}$~\cite{zhang2025synthesis}, which exhibits a superconducting transition temperature of $T_c \approx 112$~K at 215~GPa. In this system, partial substitution on the metal sublattice leads to a substantial enhancement of electron--phonon coupling and an increase of $T_c$ by approximately 60\% compared to the binary parent compound LuH$_{5.75}$. These results highlight the strong sensitivity of A15 hydrides to chemical substitution and demonstrate that controlled alloying can simultaneously modify lattice dynamics, hydrogen bonding, and electronic structure to enhance superconductivity.

Motivated by these developments, we systematically investigate the structural stability, lattice dynamics, and superconducting properties of A15-type LaH$_{5.75}$ and its Ba-alloyed variants under pressure using first-principles calculations. We find that pristine LaH$_{5.75}$ exhibits pronounced dynamical instabilities manifested as soft acoustic phonons at the $M$ point, indicating a tendency toward structural distortion. Importantly, we show that partial substitution of La by Ba stabilizes the A15 lattice, with La$_{0.75}$Ba$_{0.25}$H$_{5.75}$ emerging as a dynamically stable phase over a broad pressure range. The stabilized Ba--La alloy displays significantly enhanced electron--phonon coupling, leading to predicted superconducting transition temperatures that are approximately 94\% higher than those of pristine A15 LaH$_{5.75}$ at similar pressures. These results demonstrate that Ba--La alloying provides an effective route to stabilize A15 hydrides with reduced hydrogen content while simultaneously enhancing superconductivity. Furthermore, a Ba--La cuprate had played a major role in the history of the so-called unconventional superconductors \cite{BednorzMuller}. By linking hydrogen-lean compositions with the emerging alloy-design paradigm of Ba--La in A15 hydride superconductors, this study offers a unified framework for the rational design of new high-$T_c$ hydrides at experimentally accessible pressures. 

\section{Computational Methods}

All calculations were performed within density functional theory (DFT) using the \textsc{Quantum ESPRESSO} package \cite{giannozzi2009quantum,giannozzi2017advanced}. We implemented both the generalized gradient approximation using the Perdew--Burke--Ernzerhof (PBE)~\cite{perdew1996generalized} functional and the local density approximation (LDA), together with norm-conserving pseudopotentials constructed within the Hartwigsen-Goedecker-Hutter (HGH) scheme. Pristine A15-LaH$_{5.75}$ and Ba-alloyed La$_{1-x}$Ba$_x$H$_{5.75}$ structures were optimized at fixed pressures using the Broyden--Fletcher--Goldfarb--Shanno (BFGS) algorithm~\cite{liu1989limited}. Both lattice parameters and atomic positions were fully relaxed until the residual Hellmann--Feynman forces on all atoms were smaller than $10^{-5}$~eV/\AA.

For total-energy and formation enthalpy ($\Delta H$) calculations, the self-consistent field (SCF) electronic wave functions were expanded in a plane-wave basis set with a kinetic-energy cutoff of 100~Ry, and the charge-density cutoff was set to four times the wave-function cutoff (400~Ry). Brillouin-zone integrations were performed using a $12 \times 12 \times 12$ Monkhorst--Pack~\cite{monkhorst1976special} $k$-point mesh together with the Methfessel--Paxton smearing scheme~\cite{methfessel1989high} and a smearing width of 0.02~Ry. These parameters ensure well-converged total energies required for reliable thermodynamic comparisons.

Lattice dynamical properties and electron--phonon interactions were evaluated within density functional perturbation theory (DFPT) \cite{baroni2001phonons}. For electron--phonon coupling (EPC) calculations, a reduced plane-wave kinetic-energy cutoff of 60~Ry and a corresponding charge-density cutoff of 240~Ry were employed, which provide a good balance between numerical accuracy and computational efficiency. Brillouin-zone integrations were carried out using an $16\times16\times16$ and $8\times8\times8$ Monkhorst--Pack k-mesh.
Phonon dispersions were computed on a uniform $4\times4\times4$ $\mathbf{q}$-point mesh and interpolated along high-symmetry directions of the Brillouin zone.
Dynamical stability was assessed by inspecting the phonon spectra for imaginary frequencies.

Within DFPT, the phonon linewidth associated with phonon mode $\nu$ at wave vector $\mathbf{q}$ is given by
\begin{equation}
\gamma_{\mathbf{q}\nu}
=
2\pi \omega_{\mathbf{q}\nu}
\sum_{mn}
\sum_{\mathbf{k}}
\left|
g_{\mathbf{k}+\mathbf{q},\mathbf{k}}^{\mathbf{q}\nu,mn}
\right|^{2}
\delta(\epsilon_{\mathbf{k}+\mathbf{q},m}-\epsilon_{F})
\delta(\epsilon_{\mathbf{k},n}-\epsilon_{F}),
\end{equation}
where $g_{\mathbf{k}+\mathbf{q},\mathbf{k}}^{\mathbf{q}\nu,mn}$ is the electron--phonon matrix element, $\omega_{\mathbf{q}\nu}$ is the phonon frequency, and $\epsilon_{F}$ is the Fermi energy.

The Eliashberg spectral function $\alpha^{2}F(\omega)$ was constructed as
\begin{equation}
\alpha^{2}F(\omega)
=
\frac{1}{2N(\epsilon_{F})}
\sum_{\mathbf{q}\nu}
\delta(\omega-\omega_{\mathbf{q}\nu})
\frac{\gamma_{\mathbf{q}\nu}}{\omega_{\mathbf{q}\nu}},
\end{equation}
where $N(\epsilon_{F})$ is the electronic density of states at the Fermi level.
The total electron--phonon coupling constant $\lambda$ was obtained from $\alpha^{2}F(\omega)$ via
\begin{equation}
\lambda
=
2
\int_{0}^{\omega_{\mathrm{max}}}
\frac{\alpha^{2}F(\Omega)}{\Omega}
\, d\Omega .
\end{equation}

The logarithmic average phonon frequency was evaluated as
\begin{equation}
\omega_{\mathrm{ln}}
=
\exp\!\left[
\frac{2}{\lambda}
\int_{0}^{\infty}
\ln(\Omega)
\frac{\alpha^{2}F(\Omega)}{\Omega}
\, d\Omega
\right],
\end{equation}
and the square root of the second moment of the phonon spectrum is defined by
\begin{equation}
\omega_2
=
\left[
\frac{2}{\lambda}
\int_{0}^{\infty}
\Omega\,\alpha^{2}F(\Omega)\,d\Omega
\right]^{1/2}.
\end{equation}

The superconducting transition temperature $T_c$ was estimated using the Allen--Dynes modification of the McMillan equation~ \cite{allen1975transition,pinsook2024analytic},
\begin{equation}
T_c
=
\frac{f_1 f_2\,\omega_{\mathrm{ln}}}{1.20}
\exp\!\left[
-\frac{1.04(1+\lambda)}
{\lambda-\mu^{\ast}(1+0.62\lambda)}
\right],
\end{equation}
where $\mu^{\ast}$ is the screened Coulomb pseudopotential.
Strong-coupling and spectral-shape corrections are included through
\begin{equation}
f_1
=
\left[
1+
\left(
\frac{\lambda}
{2.46\left(1+3.8\mu^{\ast}\right)}
\right)^{3/2}
\right]^{1/3},
\end{equation}
and
\begin{equation}
f_2
=
1+
\frac{\lambda^2\left(\omega_2/\omega_{\mathrm{ln}}-1\right)}
{\lambda^2+3.3124\left(1+6.3\mu^{\ast}\right)^2}.
\end{equation}

Throughout this work, the Coulomb pseudopotential was fixed at $\mu^{\ast}=0.13$.
While the absolute values of $T_c$ depend on the choice of $\mu^{\ast}$, the relative trends discussed here are robust against moderate variations of this parameter.

\section{Results and Discussion}
    \subsection{Crystal Structure}
    \begin{figure}[h!]
    \centering
    \includegraphics[width=14cm]{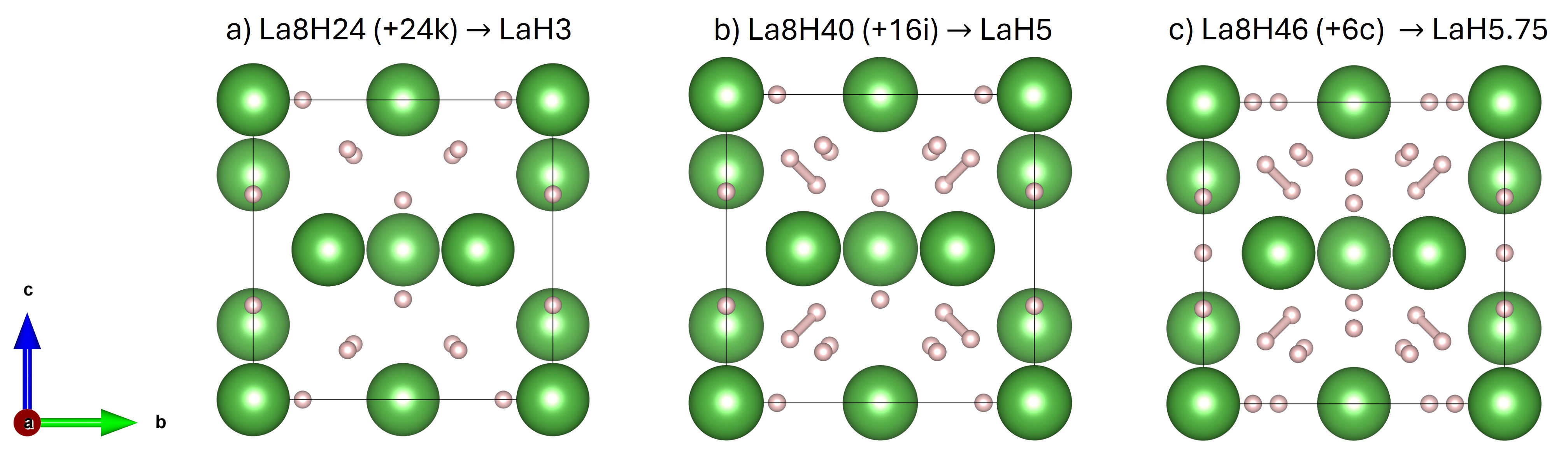}
    \caption{Crystal structures of A15-type La--H compounds in the cubic $Pm\bar{3}m$ space group, illustrating the sequential occupation of hydrogen Wyckoff sites. 
    (a) La$_8$H$_{24}$, where hydrogen atoms occupy the $24k$ sites, corresponding to the LaH$_3$ stoichiometry. 
    (b) La$_8$H$_{40}$, obtained by additional occupation of the $16i$ sites, yielding LaH$_5$. 
    (c) La$_8$H$_{46}$, where further filling of the $6c$ sites leads to the LaH$_{5.75}$ composition. 
    Green spheres represent La atoms and pink spheres denote H atoms; the cubic unit cell is shown for clarity.}
    \label{fig:a15}
\end{figure}

    The A15-type LaH$_{5.75}$ crystallizes in the cubic $Pm\bar{3}m$ space group as shown in Figure~\ref{fig:a15}. Hydrogen atoms occupying the $24k$ Wyckoff sites form LaH$_3$, while additional hydrogen atoms at the $16i$ sites give rise to LaH$_5$. Further occupation of the $6c$ Wyckoff sites leads to the LaH$_{5.75}$ stoichiometry. 

    \begin{figure}[h!]
    \centering
    \includegraphics[width=10cm]{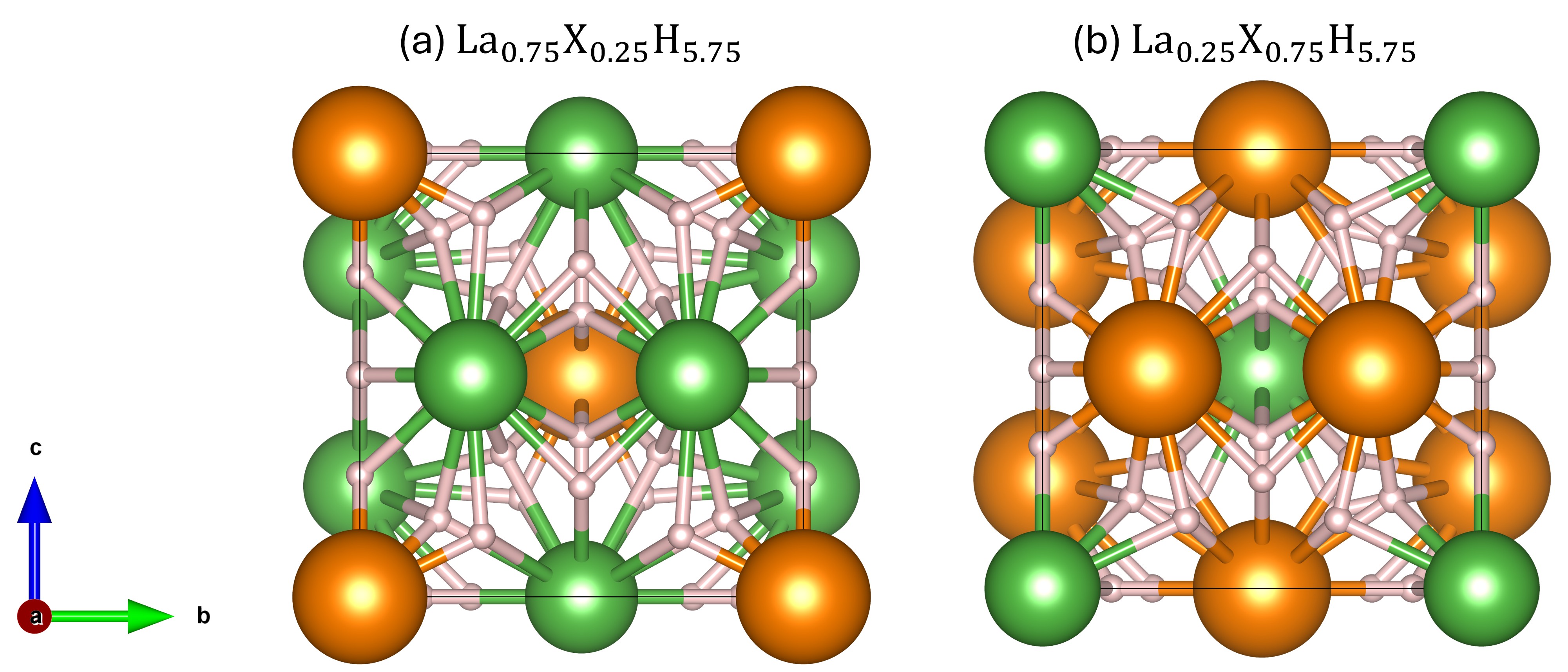}
    \caption{Crystal structures of A15-type La$_{1-x}X_{x}$H$_{5.75}$ alloys in the cubic $Pm\bar{3}m$ symmetry, where $X = \mathrm{Ba}$ or $\mathrm{Hf}$. (a) La$_{0.75}X_{0.25}$H$_{5.75}$ and (b) La$_{0.25}X_{0.75}$H$_{5.75}$, illustrating two representative substitution levels on the La sublattice. Green and orange spheres denote La and $X$ atoms, respectively, while pink spheres represent hydrogen atoms forming the A15 hydrogen framework. The crystallographic axes are indicated for reference.}
    \label{fig:x_la_alloy}
    \end{figure}

    To investigate the effect of alloying on the structural stability and symmetry of the A15 hydride framework, partial substitution of La by isovalent elements $X$ ($X = \mathrm{Ba}, \mathrm{Hf}$) was considered, forming ordered alloy compositions $X_{0.25}$La$_{0.75}$H$_{5.75}$ and $X_{0.75}$La$_{0.25}$H$_{5.75}$ as shown in Figure~\ref{fig:x_la_alloy}. While these substitutions are constructed to preserve the underlying A15 crystal structure and the cubic $Pm\bar{3}m$ symmetry, dynamical stability is found to be strongly composition dependent. In particular, only the Ba$_{0.25}$La$_{0.75}$H$_{5.75}$ alloy is dynamically stable, exhibiting no imaginary phonon modes over the pressure range of 120--200~GPa. In contrast, the other alloy compositions considered develop dynamical instabilities, indicating that the hydrogen sublattice is sensitive to both the chemical identity and concentration of the substituted cation. These results suggest that limited Ba substitution provides a viable symmetry-preserving route for tuning the electronic and vibrational properties of A15-type hydrides under high pressure. The lattice parameters of pristine and alloyed A15-type LaH$_{5.75}$ are summarized in Table~\ref{tab:lattice}.
    
    \begin{table}[h!]
\centering
\caption{Pressure dependence of the cubic lattice parameter (in \AA) for A15-type LaH$_{5.75}$ and La$_{0.75}$Ba$_{0.25}$H$_{5.75}$.}\label{tab:lattice}
\label{tab:lattice_pressure}
\begin{tabular}{|c|c|c|c|}
\hline
Pressure (GPa) & LaH$_{5.75}$ (\AA) & Pressure (GPa) & La$_{0.75}$Ba$_{0.25}$H$_{5.75}$ (\AA) \\
\hline
100 & 6.91 & 100 & 6.88 \\
110 & 6.89 & 110 & 6.85 \\
120 & 6.87 & 120 & 6.83 \\
130 & 6.85 & 130 & 6.81 \\
140 & 6.84 & 140 & 6.79 \\
150 & 6.82 & 150 & 6.77 \\
160 & 6.81 & 160 & 6.75 \\
170 & 6.79 & 170 & 6.73 \\
180 & 6.78 & 180 & 6.71 \\
190 & 6.76 & 190 & 6.69 \\
200 & 6.75 & 200 & 6.67 \\
\hline
\end{tabular}
\end{table}
    
    \begin{table}[h!]
\centering
\caption{Molecular behaviour of compounds.  Shortest bondlengths and number of H-H pair with bondlength less than 1\AA. Frequencies associated with those bond vibrations.  We note that MD runs in LaH$_{5.75}$ show that the lifetime of these molecules is very short, less than a picosecond at 300K.   }
\begin{tabular}{|c|c|c|c|c|c|}
\hline
Pressure & Functional & LaH$_{5.75}$ (\AA) & La$_{0.75}$Ba$_{0.25}$H$_{5.75}$ (\AA) & LaH$_{5.75}$ (meV) & La$_{0.75}$Ba$_{0.25}$H$_{5.75}$ (meV) \\
\hline
100 GPa & PBE & 0.7-0.8 (7) & 240-250 & 0.8-0.9 (5) & 200-220 \\

\hline
\end{tabular}
\end{table}

    \subsection{Phonons and Exchange--correlation Dependence}
    \begin{figure}[h!]
    \centering
    \includegraphics[width=15cm]{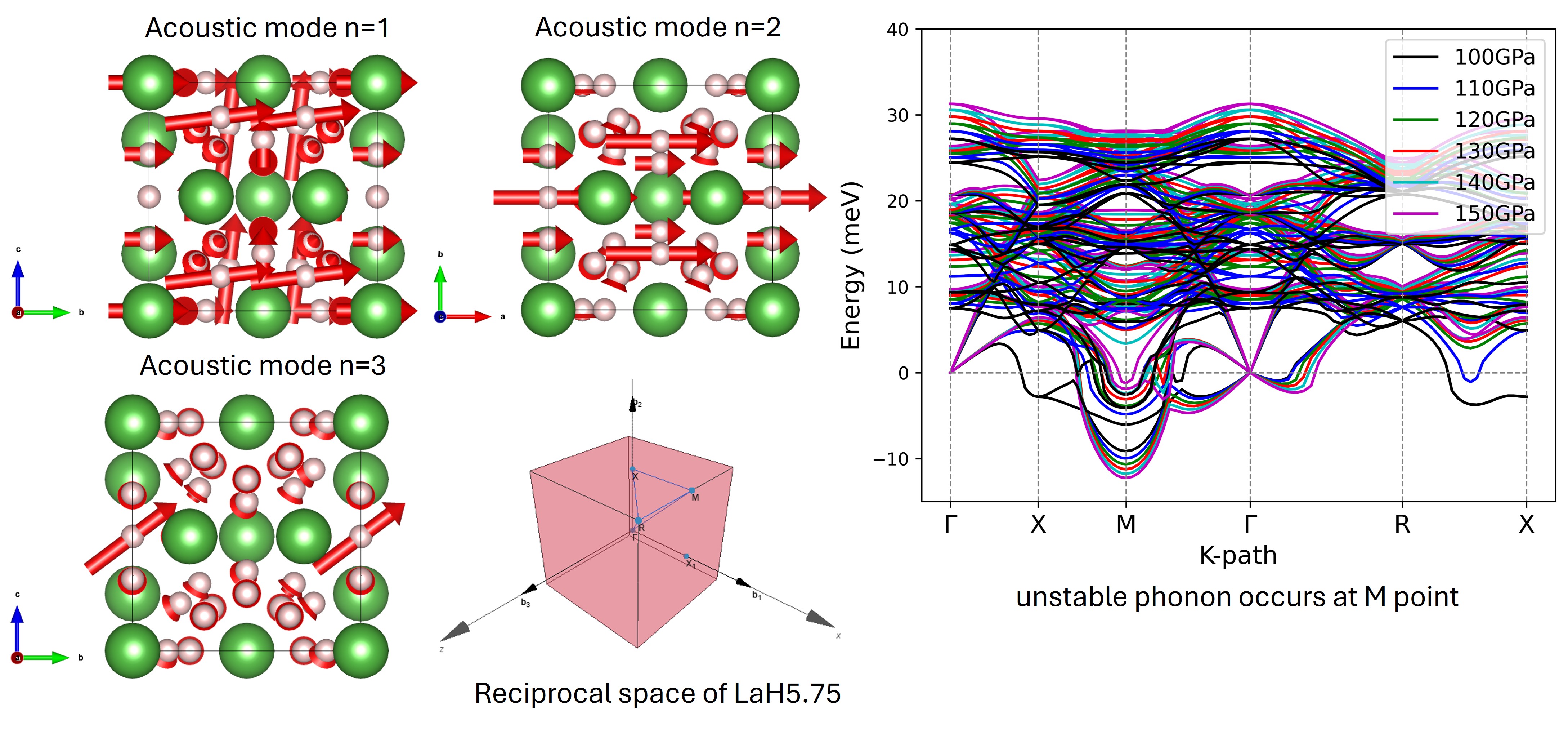}
    \caption{Phonon properties of A15-type LaH$_{5.75}$ calculated using the PBE exchange--correlation functional. The right panel shows the phonon dispersion relations along the high-symmetry path $\Gamma$--X--M--$\Gamma$--R--X for pressures ranging from 100 to 150~GPa. Imaginary phonon frequencies appear at the $M$ point, indicating a dynamical instability of the cubic $Pm\bar{3}m$ structure within PBE. The left panels display the eigenvector patterns of the three acoustic modes associated with the instability, revealing collective lattice distortions involving both La and H atoms. The Brillouin zone and the corresponding high-symmetry points of LaH$_{5.75}$ are shown for reference.}
    \label{fig:phonon_pbe}
    \end{figure}

    Figure~\ref{fig:phonon_pbe} shows the phonon dispersion relations of A15-type LaH$_{5.75}$ calculated using the PBE exchange--correlation functional over the pressure range 100--150~GPa. Pronounced dynamical instabilities are observed, characterized by imaginary phonon frequencies at the $M$ and $X$ point of the Brillouin zone. The unstable modes originate primarily from low-energy acoustic branches, as illustrated by the eigenvector analysis of the three acoustic modes, indicating collective lattice distortions involving both La and H sublattices. Although increasing pressure progressively hardens the phonon spectrum especially at the $X$ point, the instability at the $M$ point persists throughout the investigated pressure range, suggesting that within the PBE approximation the cubic $Pm\bar{3}m$ structure of LaH$_{5.75}$ is dynamically unstable. This behavior points to a tendency toward a symmetry-lowering distortion driven by soft phonons.  Freezing in these phonons typically results in the formation of weakly-bound H$_2$ molecules. We note that such molecularization in Lanthanum Hydrides can be destabilised by temperature or zero-point energy\cite{van2023competition}, and that imaginary phonons can be eliminated by including anharmonic effects\cite{errea2020quantum}.

    \begin{figure}[h!]
    \centering
    \includegraphics[width=8cm]{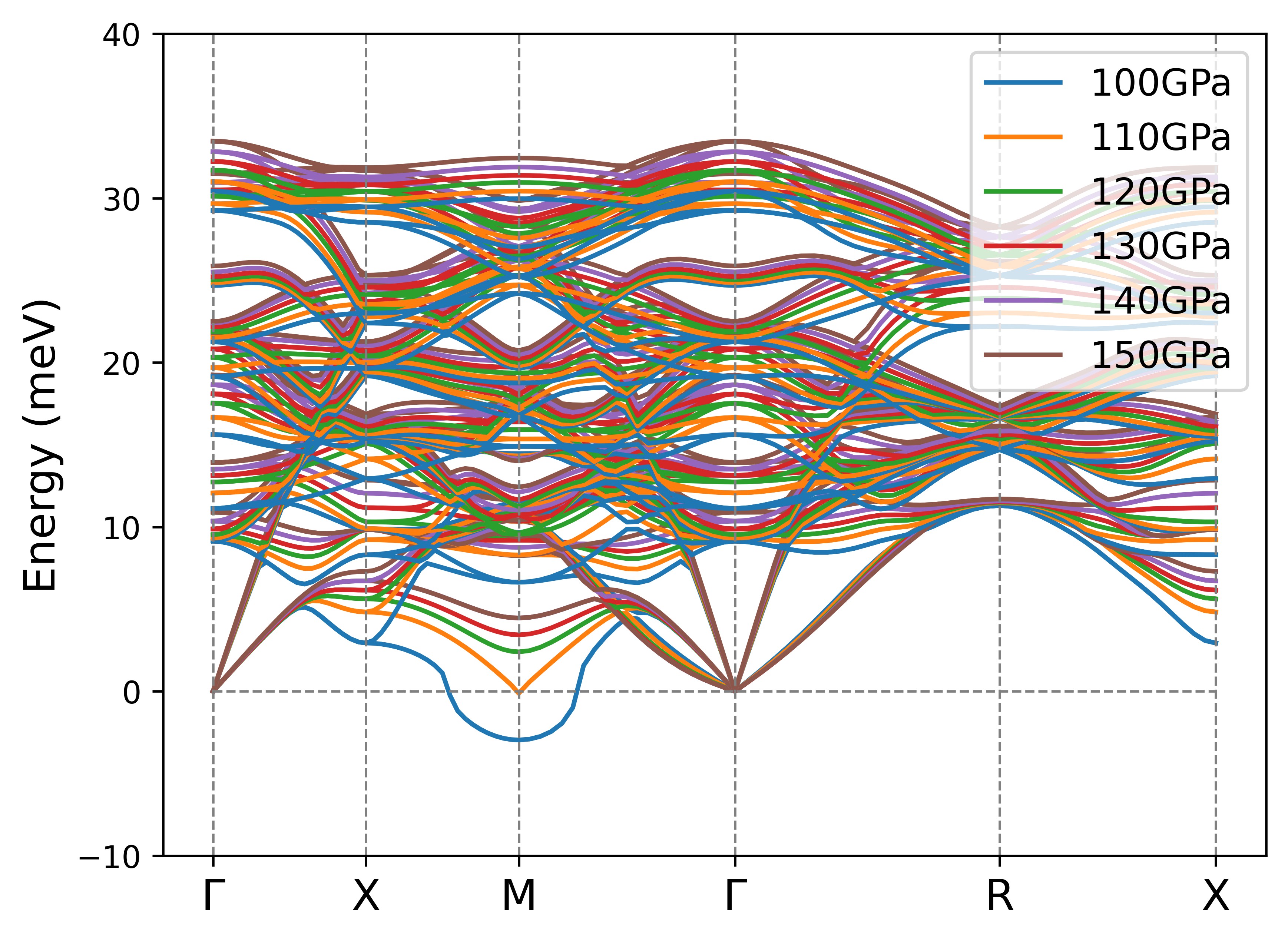}
    \caption{Phonon dispersion relations of A15-type LaH$_{5.75}$ calculated using the LDA exchange--correlation functional along the high-symmetry path $\Gamma$--X--M--$\Gamma$--R--X for pressures between 100 and 150~GPa. In contrast to the PBE results, all phonon modes remain real across the Brillouin zone for pressures above 120~GPa, indicating full dynamical stability of the cubic $Pm\bar{3}m$ structure within LDA. The hardening of the low-energy acoustic modes, particularly near the $M$ point, reflects the enhanced bonding and reduced equilibrium volume characteristic of LDA under high pressure.}
    \label{fig:phonon_lda}
    \end{figure}

    Figure~\ref{fig:phonon_lda} presents the corresponding phonon dispersions calculated using the LDA exchange--correlation functional. In contrast to the PBE results, all phonon modes remain real across the entire Brillouin zone for pressures above 120~GPa, indicating full dynamical stability of the A15 structure within LDA. The previously unstable acoustic modes at the $M$ point are significantly hardened, reflecting the enhanced bonding and reduced equilibrium volume characteristic of LDA. These results demonstrate a strong exchange-correlation dependence of the lattice dynamics in LaH$_{5.75}$, highlighting the sensitivity of the dynamical stability to subtle changes in the electronic structure and equilibrium geometry under high pressure. Therefore, throughout the paper, all calculations are performed using the LDA exchange--correlation functional.

    The use of LDA as a reference functional is physically well motivated in the present high-pressure regime. At pressures exceeding 100~GPa, the electronic density is strongly enhanced and interatomic distances are significantly reduced, conditions under which the local-density approximation is known to provide an accurate description of short-range exchange--correlation effects. In hydrogen-rich systems, lattice dynamics are extremely sensitive to equilibrium volume and bonding strength due to the light mass and large vibrational amplitudes of hydrogen. Compared to PBE, LDA predicts smaller equilibrium lattice parameters and stronger La--H and H--H interactions, which increase lattice restoring forces and harden low-energy acoustic phonon modes that are unstable within PBE. Although anharmonic effects are expected to be significant in such systems, the enhanced lattice stiffness captured by LDA at the harmonic level leads to a phonon renormalization qualitatively similar to the stabilizing effect of anharmonicity. Notably, LDA yields a dynamically stable phonon spectrum and a superconducting transition temperature in close agreement with experimental observations for A15-type LaH$_{5.75}$~\cite{lah575_prl,cross2024high,guo2024unusual}, supporting its use as a physically meaningful reference for describing lattice stability and superconductivity under high pressure.

    \subsection{Electronic Properties}
\begin{figure}[h!]
  \centering
  \includegraphics[width=13cm]{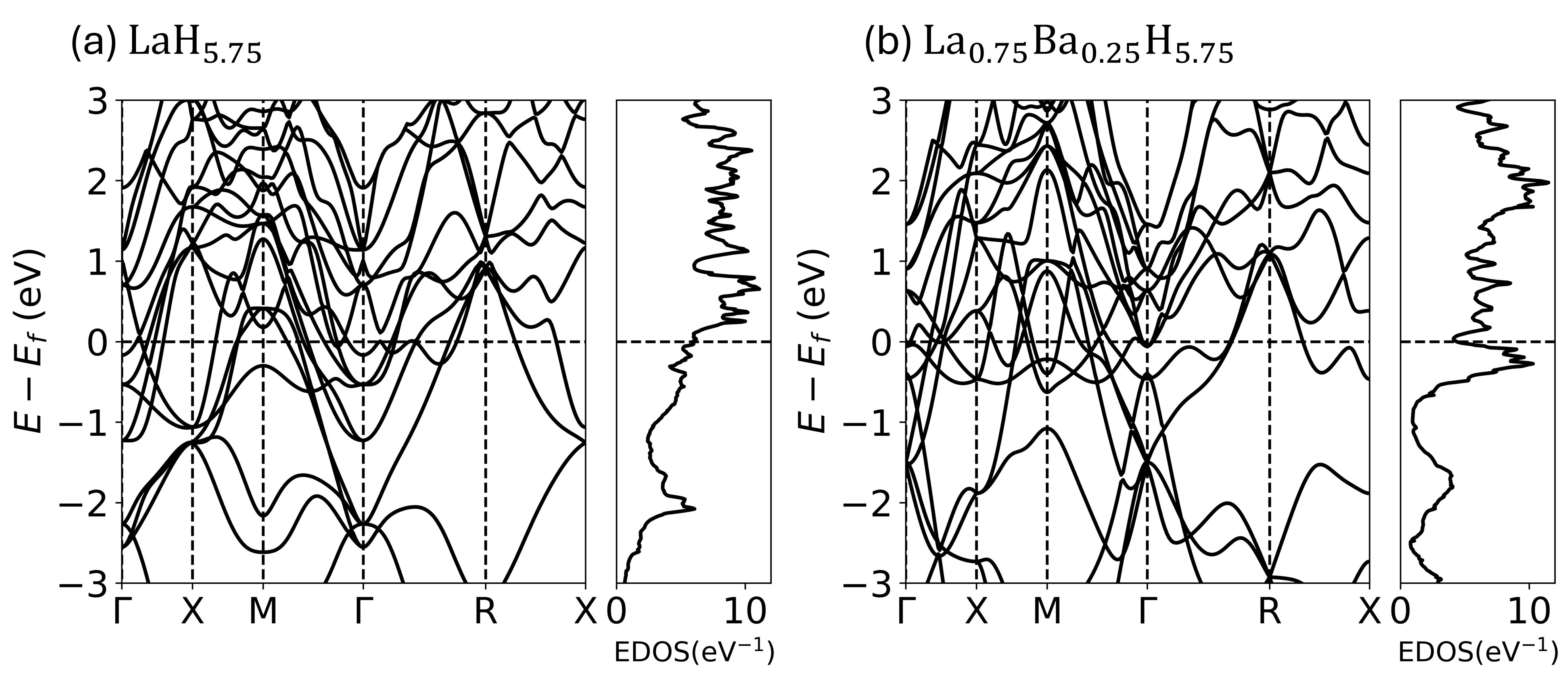}
  \caption{Electronic band structures and electronic density of states (EDOS) of (a) LaH$_{5.75}$ at 120~GPa and (b) La$_{0.75}$Ba$_{0.25}$H$_{5.75}$ at 130~GPa. The band energies are plotted along the high-symmetry path $\Gamma$--X--M--$\Gamma$--R--X of the Brillouin zone. The Fermi level $E_f$ is set to zero energy and indicated by the horizontal dashed line. The right panels display the total EDOS, highlighting a finite density of states at the Fermi level for both compounds, consistent with their metallic character.}
  \label{fig:bands_dos}
\end{figure}

The electronic band structures and electronic density of states (EDOS) of LaH$_{5.75}$ and La$_{0.75}$Ba$_{0.25}$H$_{5.75}$ are shown in Fig.~\ref{fig:bands_dos} (a) and (b), respectively. For both compounds, several bands cross the Fermi level along the high-symmetry directions $\Gamma$--X--M--$\Gamma$--R--X, clearly indicating metallic behavior. This metallicity is further confirmed by the finite EDOS at the Fermi level, which is an essential prerequisite for phonon-mediated superconductivity.

In pristine LaH$_{5.75}$ at 120~GPa, the electronic bands near the Fermi level are relatively dispersive, reflecting substantial hybridization between La-derived states and the hydrogen sublattice. The corresponding electronic density of states (EDOS) around $E_f$ is moderately smooth, indicating a well-connected Fermi surface without a pronounced pseudogap. Upon partial substitution of La by Ba, noticeable modifications of the low-energy electronic structure emerge. In La$_{0.75}$Ba$_{0.25}$H$_{5.75}$ at 130~GPa, Ba incorporation alters the band dispersion in the vicinity of $E_f$, leading to a redistribution of electronic states near the Fermi level. While the electronic density of states at $E_f$ remains comparable to that of pristine LaH$_{5.75}$, it is significantly enhanced in the energy range slightly below $E_f$, which is consistent with the observed strengthening of electron--phonon coupling discussed in the following sections.

    \subsection{Electron-Phonon Coupling and Superconductivity}

\begin{figure}[h!]
  \centering
  \includegraphics[width=11cm]{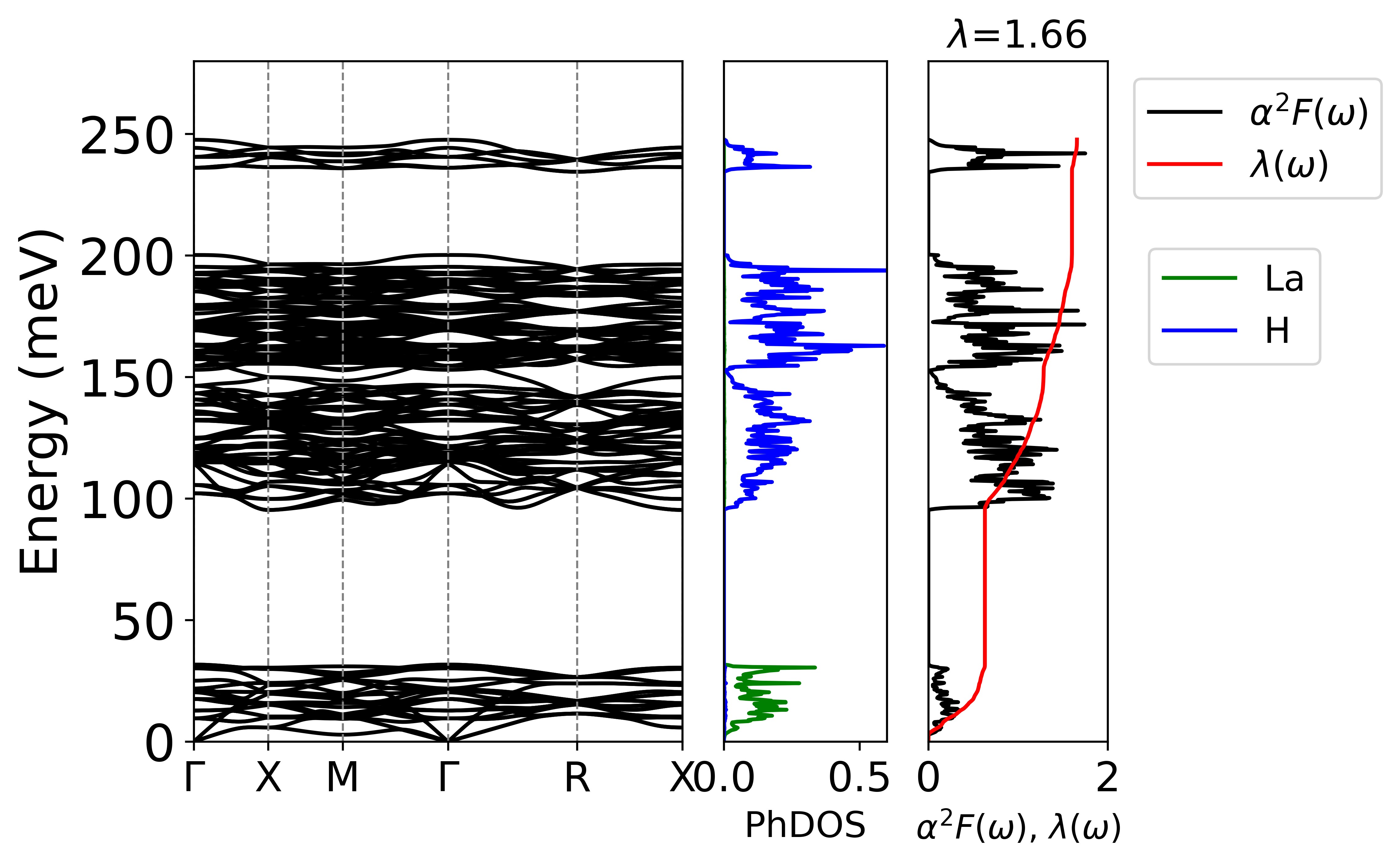}
  \caption{Phonon dispersion relations, phonon density of states (PhDOS), Eliashberg spectral function $\alpha^2F(\omega)$, and the cumulative electron--phonon coupling constant $\lambda(\omega)$ of LaH$_{5.75}$ at 120~GPa. The phonon dispersions are plotted along the high-symmetry path $\Gamma$--X--M--$\Gamma$--R--X. The absence of imaginary phonon frequencies confirms the dynamical stability of the structure. The right panels show the PhDOS, $\alpha^2F(\omega)$, and $\lambda(\omega)$, with the total electron--phonon coupling constant reaching $\lambda = 1.66$.}
  \label{fig:ph_epc_lah575}
\end{figure}

The phonon dispersion relations, phonon density of states (PhDOS), Eliashberg spectral function $\alpha^2F(\omega)$, and the integrated electron--phonon coupling constant $\lambda(\omega)$ of LaH$_{5.75}$ at 120~GPa are shown in Fig.~\ref{fig:ph_epc_lah575}. The phonon spectrum exhibits no imaginary frequencies throughout the Brillouin zone, confirming the dynamical stability of LaH$_{5.75}$ at this pressure. The vibrational spectrum can be broadly divided into low-frequency modes below $\sim$40~meV, mainly associated with La vibrations, and high-frequency modes extending from $\sim$100~meV up to $\sim$250~meV, originating from the hydrogen sublattice.

The corresponding PhDOS reveals that the dominant contribution to the high-energy phonon modes arises from hydrogen vibrations, while the low-energy part is governed by La-related modes. The Eliashberg spectral function $\alpha^2F(\omega)$ shows pronounced contributions over a wide energy range, indicating strong coupling between electrons and both low- and intermediate-frequency phonons. The cumulative electron--phonon coupling strength $\lambda(\omega)$ increases steadily with phonon energy and saturates at a total value of $\lambda = 1.66$, highlighting the strong-coupling nature of superconductivity in this compound.

Using the Allen--Dynes modified McMillan formalism, we obtain a logarithmic average phonon frequency of $\omega_{\log} = 607$~K and a second moment frequency of $\omega_2 = 1332$~K. These relatively high characteristic phonon frequencies, combined with the large electron--phonon coupling constant, result in a superconducting transition temperature of $T_c^{\mathrm{(AD)}} = 94.5$~K at 120~GPa. The strong contribution of hydrogen-derived phonon modes to $\alpha^2F(\omega)$ underscores the crucial role of the hydrogen sublattice in enhancing the electron--phonon interaction and stabilizing high-temperature superconductivity in LaH$_{5.75}$. This predicted $T_c$ is in good agreement with recent experimental observations, which report superconductivity in the type-I clathrate A15-LaH$_{5.75-x}$ with a transition temperature of approximately 98~K near 94~GPa~\cite{lah575_prl}, at 90~K near 95~GPa~\cite{cross2024high}, and 105~K near 118~GPa~\cite{guo2024unusual}. The extracted $\omega_{log} \approx$ 840K from the experiment~\cite{lah575_prl} (see Appendix A), compares with 607K from our calculation of LaH$_{5.75}$ at 120 GPa. This discrepancy could come from a slightly different $x$ fraction or pressure.

\begin{figure}[h!]
    \centering
    \includegraphics[width=12cm]{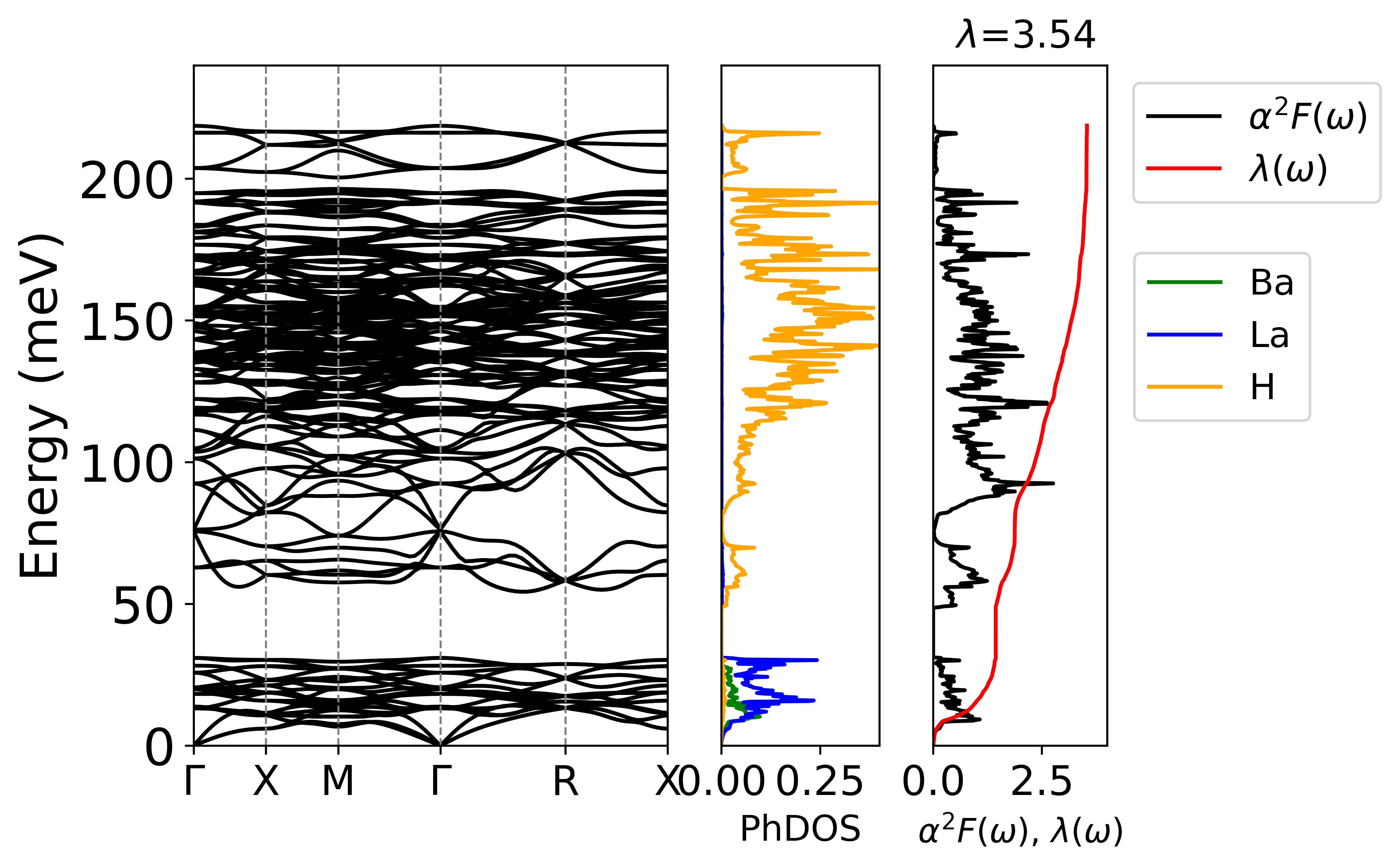}
    \caption{Phonon dispersion relations, phonon density of states (PhDOS), Eliashberg spectral function $\alpha^{2}F(\omega)$, and integrated electron--phonon coupling constant $\lambda(\omega)$ for Ba$_{0.25}$La$_{0.75}$H$_{5.75}$ at 130~GPa. The phonon spectrum exhibits a clear separation between low-frequency metal-dominated modes and high-frequency hydrogen-derived vibrations. The electron--phonon coupling is dominated by hydrogen modes, leading to a large total coupling constant $\lambda = 3.54$, with a logarithmic average phonon frequency $\omega_{\log} = 503$~K and a second moment $\omega_{2} = 1056$~K.}
    \label{fig:epc_ba025}
\end{figure}

Figure~\ref{fig:epc_ba025} presents the phonon dispersion, phonon density of states (PhDOS), Eliashberg spectral function $\alpha^2F(\omega)$, and the integrated electron--phonon coupling constant $\lambda(\omega)$ of Ba$_{0.25}$La$_{0.75}$H$_{5.75}$ at 130~GPa where Below 130~GPa, imaginary phonon modes appear, indicating dynamical instability.  The phonon spectrum is characterized by low-frequency modes dominated by Ba/La vibrations below 40~meV and high-frequency hydrogen-derived modes extending from above 50~mev to 220~meV. The separation between metal- and hydrogen-dominated vibrations results in a broad $\alpha^2F(\omega)$ distribution, with the dominant contribution to the total electron--phonon coupling arising from intermediate- and high-frequency hydrogen modes.

At 130~GPa, the total electron--phonon coupling constant reaches $\lambda = 3.54$, accompanied by a logarithmic average phonon frequency $\omega_{\log} = 503$~K and a second moment $\omega_2 = 1056$~K. These values indicate a strong-coupling superconducting regime. Using the Allen--Dynes modified McMillan formula with a standard Coulomb pseudopotential $\mu^{\ast}=0.13$, the superconducting critical temperature is estimated to be $T_c \approx 183.2$~K.

\begin{table}[h!]
\centering
\caption{Pressure dependence of the formation enthalpy difference $\Delta H$, electron--phonon coupling constant $\lambda$, logarithmic average phonon frequency $\omega_{\log}$, second moment $\omega_{2}$, and superconducting critical temperature $T_c$ for Ba$_{0.25}$La$_{0.75}$H$_{5.75}$. The values of $T_c$ are calculated using the Allen--Dynes modified McMillan formula with $\mu^{\ast}=0.13$. Below 130~GPa, the phonon spectrum exhibits dynamical instabilities.}
\label{tab:tc_ba025}
\begin{tabular}{|c|c|c|c|c|c|}
\hline
Pressure (GPa) & $\Delta H$ (eV) & $\lambda$ & $\omega_{\log}$ (K) & $\omega_{2}$ (K) & $T_c$ (K) \\
\hline
100 & -0.97  & --   & --  & --   & --   \\
110 & -0.91  & --   & --  & --   & --   \\
120 & -0.86  & --   & --  & --   & --   \\
130 & -0.81  & 3.54 & 503 & 1056 & 183.2 \\
140 & -0.75  & 2.90 & 644 & 1171 & 175.6 \\
150 & -0.70  & 2.66 & 712 & 1234 & 173.3 \\
160 & -0.64 & 2.46 & 774 & 1297 & 171.1 \\
170 & -0.59 & 2.33 & 795 & 1341 & 166.8 \\
180 & -0.54 & 2.19 & 818 & 1384 & 160.5 \\
190 & -0.50 & 2.09 & 879 & 1429 & 161.8 \\
200 & -0.45 & 2.03 & 888 & 1464 & 159.0 \\
\hline
\end{tabular}
\end{table}

The pressure dependence of the formation enthalpy difference,
\[
\Delta H = H_{\mathrm{Ba_{0.25}La_{0.75}H_{5.75}}} - \left[0.25\,H_{\mathrm{BaH_{5.75}}} + 0.75\,H_{\mathrm{LaH_{5.75}}}\right],
\]
together with the calculated superconducting parameters, is summarized in Table~\ref{tab:tc_ba025} where H is energy from DFT static optimization. Over the entire pressure range investigated (100--200~GPa), the formation enthalpy $\Delta H$ remains negative and decreases monotonically with increasing pressure, indicating that Ba$_{0.25}$La$_{0.75}$H$_{5.75}$ is thermodynamically favored relative to phase separation based on static-lattice structural optimizations. Although $\Delta H$ remains negative over the entire pressure range, it becomes less negative with increasing pressure, indicating a gradual reduction of the thermodynamic driving force for alloy formation under compression. Concurrently, the electron--phonon coupling strength $\lambda$ decreases gradually with increasing pressure, while the characteristic phonon frequencies $\omega_{\log}$ and $\omega_2$ increase due to lattice stiffening. As a result, the superconducting transition temperature exhibits a moderate reduction, decreasing from $T_c \approx 183$~K at 130~GPa to $T_c \approx 159$~K at 200~GPa, while remaining firmly within the strong-coupling regime throughout the entire pressure range considered.

We note that the reported formation enthalpies do not include zero-point energy (ZPE) contributions, which can be significant in hydrogen-rich systems. Previous studies have shown that the inclusion of ZPE generally stabilizes structures~\cite{mcmahon2011high,seeyangnok2025solid,seeyangnok2025hydrogenation}. Consequently, ZPE is expected to further enhance the thermodynamic stability of Ba$_{0.25}$La$_{0.75}$H$_{5.75}$ and may strengthen its energetic preference relative to competing phases. Although quantum nuclear effects are not explicitly accounted for here, the present results already establish Ba$_{0.25}$La$_{0.75}$H$_{5.75}$ as a dynamically stable, strong-coupling superconductor over a wide pressure range, highlighting alloying as an effective strategy for tuning superconductivity in A15-type hydrides.
\section{Discussion}

High pressure superconductivity in hydrogen-rich materials requires destabilising  H$_2$ molecular units to create atomic, metallic hydrogen.
The A-15 structure is topologically close packed, which means it has tetragonal interstitial sites large enough to contain a single hydrogen atom.  There are three distinct sites, labelled 24k 16i and 6c.   Although it is still possible to form H$_2$ units with hydrogens at adjacent sites, the bond goes "through" a triangle of heavy atoms and is therefore destablised.  High pressure, temperature, and zero-point motion all contribute further destablise the molecules.

Calculations on A-15 Lanthanum hydrides show that the superconductivity is linked to occupation of the 6c sites.
Unfortunately, the 6c sites are the least favoured for hydrogen occupation and are depopulated on depressurization.   Filling the 24k sites only leads to an ionic LaH$_3$ compound, while the 16i sites are prone to molecule formation.

Therefore, material design for superconductivity using ternary alloying is aimed at stabilising the 6c sites and destabilising the molecule formation.  Here, we showed that replacing La3+ ions with Ba2+ increases the superconductivity by a factor of 2, primarily due to increasing the electron-phonon coupling parameter $\lambda$ by a similar amount.

\section{Conclusions}
In this work, we have systematically investigated the lattice stability, electronic structure, and superconducting properties of hydrogen-deficient A15-type LaH$_{5.75}$ and its Ba-alloyed variants under high pressure using first-principles density functional theory and density-functional perturbation theory. The calculation yields a superconducting transition temperature of $T_c \approx 94.5$~K at 120~GPa, in good agreement with experimental reports of $T_c \approx 98$~K at 94~GPa, $T_c \approx 90$~K at 95~GPa, and $T_c \approx 105$~K near 118~GPa. We demonstrate that partial substitution of La by Ba provides an effective route to stabilize the A15 framework. Among the alloy compositions considered, La$_{0.75}$Ba$_{0.25}$H$_{5.75}$ emerges as the only dynamically stable phase, maintaining the cubic $Pm\bar{3}m$ symmetry over a broad pressure range from 120 to 200~GPa. The stabilization is accompanied by substantial modifications of the phonon spectrum and electronic states near the Fermi level, leading to a pronounced enhancement of the electron--phonon coupling.

Thermodynamic analysis based on static-lattice formation enthalpies indicates that La$_{0.75}$Ba$_{0.25}$H$_{5.75}$ is energetically favored with respect to phase separation over the entire pressure range considered. While zero-point energy contributions are not included in the present study, their incorporation is expected to further stabilize the alloyed phase. This suggests that Ba-alloyed A15 hydrides could be experimentally accessible at pressures comparable to those already achieved for binary A15 lanthanum hydrides, and exhibit significantly higher superconducting temperatures.

A detailed analysis of the Eliashberg spectral function shows that superconductivity in the thermodynamically stabilized alloy La$_{0.75}$Ba$_{0.25}$H$_{5.75}$ is dominated by hydrogen-derived vibrational modes, placing the system in a strong-coupling regime with electron--phonon coupling constants reaching $\lambda \approx 3.5$ at 130~GPa, corresponding to a superconducting transition temperature of approximately 183~K. The predicted superconducting transition temperature remains above 160~K up to 200~GPa, with electron--phonon coupling constants exceeding $\lambda \approx 2$, placing this compound among the highest-$T_c$ hydrogen-lean hydrides reported to date.

In general, our results establish Ba--La alloying as a powerful strategy to simultaneously stabilize hydrogen-deficient A15 hydrides and enhance their superconducting properties. This work extends the emerging alloy-design paradigm in A15 hydride superconductors and provides concrete guidance for the experimental exploration of new high-$T_c$ phases at reduced hydrogen content and experimentally relevant pressures.

    \section*{Data Availability}
    The data that support the findings of this study are available from the corresponding
    authors upon reasonable request.
    
    \section*{Code Availability}
    The first-principles DFT calculations were performed using the open-source Quantum ESPRESSO package, available at \url{https://www.quantum-espresso.org}, along with pseudopotentials from the Quantum ESPRESSO pseudopotential library at \url{https://pseudopotentials.quantum-espresso.org/}.
\section*{Acknowledgments}
	This research project is supported by the Second Century Fund (C2F), Chulalongkorn University. We acknowledge the supporting computing infrastructure provided by NSTDA, CU, CUAASC, NSRF via PMUB [B05F650021, B37G660013] (Thailand). (\url{URL:www.e-science.in.th}). This also work used the ARCHER2 UK National Supercomputing Service (\url{https://www.archer2.ac.uk}) as part of the UKCP collaboration.

    \section*{Author Contributions}
    Jakkapat Seeyangnok performed all calculations, analyzed the results, wrote the initial draft of the manuscript, and coordinated the project. Graeme J. Ackland and Udomsilp Pinsook contributed to the analysis of the results and to the writing and revision of the manuscript.

\section*{Appendix A}
\begin{figure}[h!]
    \centering
    \includegraphics[width=12cm]{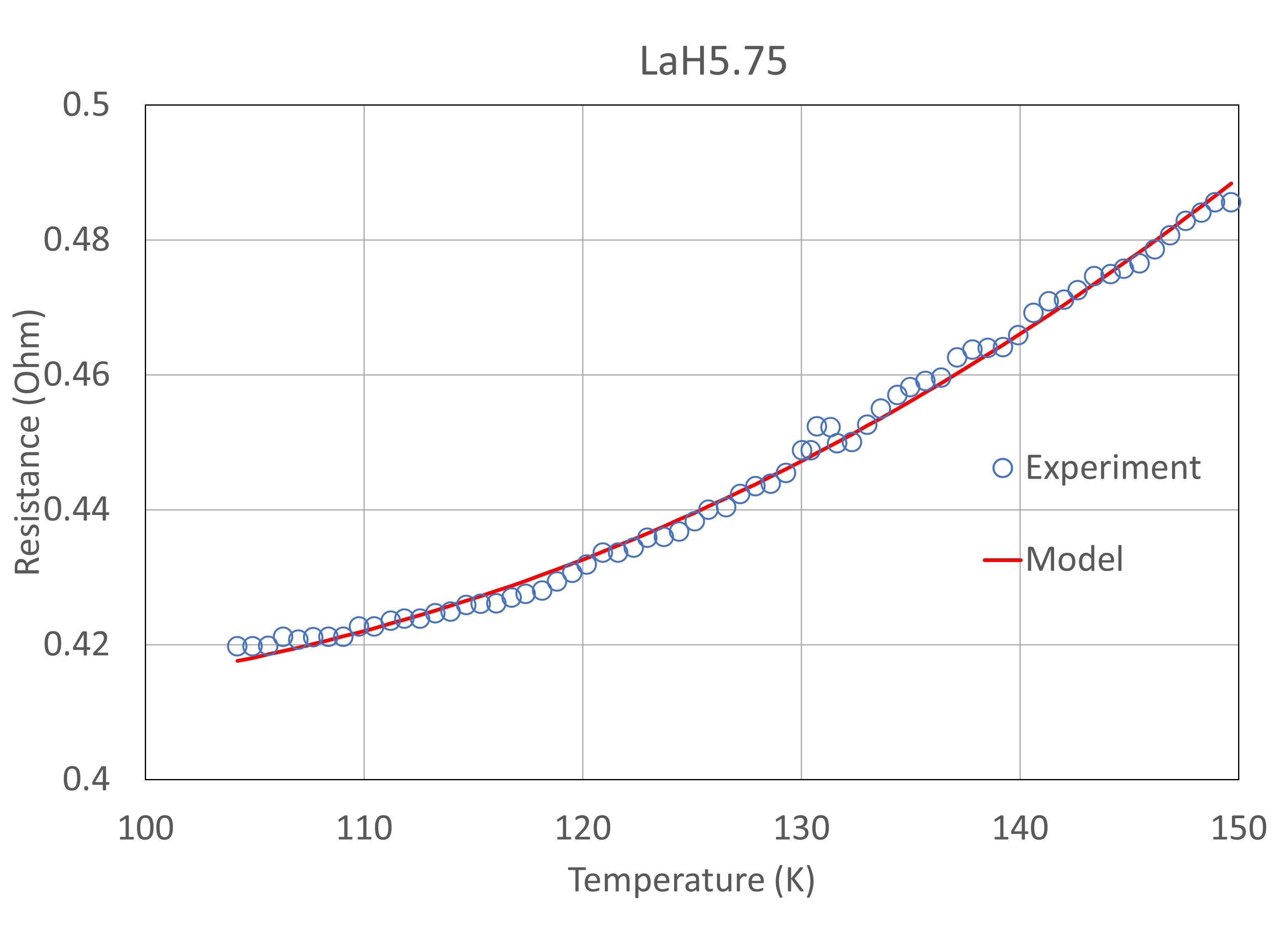}
    \caption{The resistance of LaH$_{5.75-x}$ at 98 GPa (opened blue circles)\cite{lah575_prl}, compared with the model (red line)\cite{JRCooper,UniversalResistivity}. The extracted $\omega_{log} \approx$ 840K, compared with 607K from our calculation of LaH$_{5.75}$ at 120 GPa.}
    \label{Resistance}
\end{figure}

In a solid with dominated optical phonons, the resistance can be written in the form \cite{JRCooper, UniversalResistivity}
\[
R(T) = R_0 + A \frac{\theta^2_E}{T} \frac{1}{(e^{\frac{\theta_E}{T}}-1)(1-e^{\frac{-\theta_E}{T}})},
\]
where $R_0$ is a residue resistance, $A$ is a constant, $\theta_E$ is the Einstein frequency. By fitting to an experimental result, the parameters $R_0$, $A$ and $\theta_E$ can be evaluated.   $R_0$ and $A$ are scaling parameters which relate the fundamental property, resistivity, to the measured quantity, resistance, and so  will depend on the geometry of the sample.   $\theta_E$ is a fundamental material property: Pinsook and Tanthum showed that $\theta_E$ is a first-order approximation to $\omega_{log}$ \cite{UniversalResistivity}. From Fig.~\ref{Resistance}, it shows the measured resistance of LaH$_{5.75-x}$ at 98 GPa\cite{lah575_prl}. The extracted $\omega_{log} \approx$ 840K, compared with 607K from our calculation of LaH$_{5.75}$ at 120 GPa. The discrepancy could come from higher order effects, a slightly different $x$ fraction of from different pressures.

\bibliographystyle{unsrt}
\bibliography{references}

\end{document}